\documentclass[preprint,12pt]{elsarticle}
\usepackage{amssymb}
\usepackage{amsmath}

\journal{Physics Letters B}

\graphicspath{{figures/}}

\begin{document}

%		setting options for natbib
%\setcitestyle{numbers,square,comma}
\biboptions{numbers,square,comma,merge}

\begin{frontmatter}

\title{Observable Gravity Waves From Supersymmetric Hybrid Inflation}

\author{Qaisar Shafi}
\author{Joshua R. Wickman\corref{jrw}}
\cortext[jrw]{Corresponding author}%{Email: jwickman@udel.edu}
\ead{jwickman@udel.edu}
\address{Bartol Research Institute, Department of Physics and Astronomy, 
University of Delaware, Newark, Delaware 19716, USA}

\begin{abstract}

We identify models of supersymmetric hybrid inflation in which the tensor-to-scalar ratio, a canonical measure of gravity waves produced during inflation, can be as large as 0.03 or so,   which will be tested by the Planck satellite experiment. The scalar spectral index lies within the WMAP one sigma bounds, while $|d n_s / d\ln k| \lesssim 0.01$.

\end{abstract}

\end{frontmatter}

%\label{intro}
\section*{Introduction}

Supersymmetric (SUSY) models of hybrid inflation~\citep{Dvali:1994ms,Copeland:1994vg,Linde:1993cn,Linde:1997sj,Lazarides:2001zd,Lyth:1998xn,*Mazumdar:2010sa} are enticing due to their ties to grand unified theories (GUTs) and mainstream particle physics.  In addition to utilizing the pervasive framework of SUSY, these models naturally incorporate the breaking of a gauge group $G$ into the inflationary mechanism, whether at the end of inflation (standard scenario) or during inflation (shifted scenario~\citep{Jeannerot:2000sv,Senoguz:2003zw}).  The gauge symmetry $G$ may be associated with a GUT, such as flipped SU(5)~\citep{Kyae:2005nv}, or perhaps an extension of the Standard Model (e.g. an extra U(1)$_{B-L}$ symmetry).  With the recent launch of the Planck satellite, and in anticipation of the new results it promises to yield, a thorough exploration of the predictions of such highly motivated models may bring us closer to understanding the nature of inflation.

Primordial gravitational waves in the inflationary epoch can source tensor fluctuations in the cosmic microwave background (CMB).  A convenient parametrization of these tensor modes is obtained by comparison to the scalar perturbations,
\begin{equation}
r \equiv \frac{\Delta_t^2}{\Delta_s^2}.
\end{equation}
This `tensor-to-scalar ratio' is observably large only if the inflaton field $\phi$ changes amplitude over an interval near to or greater than the Planck scale, owing to the Lyth bound~\citep{Lyth:1996im,Lyth:2009zz}
\begin{equation}
r \lesssim 0.08 \left( \frac{\Delta\phi}{m_P} \right)^2,
\label{lyth1}
\end{equation}
where $m_P \approx 2.4 \times 10^{18}$~GeV is the reduced Planck mass.  This constraint most readily allows sizable $r$-values if $\Delta\phi \gg m_P$.  Recently, some progress has been made in devising realistic inflationary models having observable $r$~($\sim 0.01$ or so) in which the inflaton amplitude is sub-Planckian~\citep{Rehman:2009wv}.  Indeed, we will require that the field amplitude remain below the Planck scale in order to ensure that supergravity corrections remain under control.

Thus far, models of SUSY hybrid inflation have predicted relatively tiny values of $r$~($\lesssim 10^{-4}$), leaving little hope of observation by current or future experimental endeavors (see, for example, Refs.~\citep{Dvali:1994ms,urRehman:2006hu,Rehman:2009nq,Rehman:2009yj}).  In this letter, we will show that the inclusion of higher order corrections from multiple sources can open up a region of parameter space that can support $r$-values large enough to be observed by the current Planck satellite observatory.

%%%%%%%%%%%%%%%%%%%%%%%%%%%%%%%%%%%%%%%%%%%%%%%%%

%\label{background}
\section*{Supersymmetric Hybrid Inflation}

Within the framework of supersymmetry, hybrid inflation is achieved via the superpotential~\citep{Dvali:1994ms,Copeland:1994vg}
\begin{equation} 
W = \kappa S(\Phi \overline{\Phi} - M^{2}) \, ,
\label{superpot}
\end{equation}
where $S$ is a gauge singlet superfield whose scalar component acts as the inflaton, and $\Phi$, $\overline{\Phi}$ are gauge conjugate superfields transforming nontrivially under some gauge group $G$.  It is often desirable to include an additional U(1) `$R$-symmetry' in SUSY theories (e.g. to curb runaway proton decay).  In this case, $S$ transforms under U(1)$_R$ in the same way as $W$, and Eq.~(\ref{superpot}) is the most general superpotential consistent with U(1)$_R$ and $G$ at the renormalizable level.

We are interested in exploring the possibility of large primordial gravity wave amplitudes produced in these models of inflation.  Consequently, it is insufficient to consider only global SUSY; supergravity (SUGRA) effects must be taken into account.  The K\"ahler potential may be written as an expansion in powers of $1/m_P$:
%\begin{equation} 
\begin{multline}
K = |S|^2 + |\Phi|^2 + |\overline{\Phi}|^2 + \frac{\kappa_S}{4}\frac{|S|^4}{m_P^2} + \frac{\kappa_\Phi}{4}\frac{|\Phi|^4}{m_P^2} +\frac{\kappa_{\overline{\Phi}}}{4}\frac{|\overline{\Phi}|^4}{m_P^2} \\
 + \kappa_{S \Phi}\frac{|S|^2|\Phi|^2}{m_P^2} + \kappa_{S \overline{\Phi}}\frac{|S|^2|\overline{\Phi}|^2}{m_P^2} + \kappa_{\Phi \overline{\Phi}}\frac{|\Phi|^2|\overline{\Phi}|^2}{m_P^2} + \frac{\kappa_{SS}}{6}\frac{|S|^6}{m_P^4} + \cdots ,
\label{kahler}
\end{multline}
%\end{equation}
where we have allowed only mod-square combinations of the fields in order that the U(1)$_R$ and $G$ symmetries remain intact.  In a number of recent analyses, it has been shown that the scalar spectral index $n_s$ can obtain the WMAP central value in SUSY models of inflation using a minimal~\citep{Rehman:2009nq,Rehman:2009yj} or non-minimal~\citep{BasteroGil:2006cm,urRehman:2006hu} form of the K\"ahler potential in Eq.~(\ref{kahler}).  The value of $n_s$ is expected to be an important consideration in determining the correct model of inflation; however, these models have predicted exceedingly small values of $r$, and will be handily ruled out should Planck measure a sizable tensor amplitude.  As we will see, both a non-minimal K\"ahler potential and a substantial inflaton soft mass are needed in order to produce large $r$-values.

The $F$-term SUGRA scalar potential is derived from Eqs.~(\ref{superpot}) and (\ref{kahler}) according to the formula
\begin{equation} 
V_{F} = e^{K/m_{P}^{2}} \left( K_{ij}^{-1}D_{z_{i}}WD_{z^{*}_j}W^{*} - 3m_{P}^{-2}\left| W\right| ^{2} \right),
\label{VF1}
\end{equation}
where $z_{i}\in \{s, \phi , \overline{\phi }, \cdots\}$ are the scalar components of the superfields $S$, $\Phi$, $\overline{\Phi}$, and where we have defined
\begin{eqnarray*}
K_{ij} &\equiv& \frac{\partial ^{2}K}{\partial z_{i}\partial z_{j}^{*}}, \\
D_{z_{i}}W &\equiv& \frac{\partial W}{\partial z_{i}}+m_{P}^{-2}\frac{\partial K}{\partial z_{i}}W, \\ %\,\,\,\,\,\, 
D_{z_{i}^{*}}W^{*} &=& \left( D_{z_{i}}W\right) ^{*}.
\end{eqnarray*}
Along the $D$-flat direction, $|\phi| = |\overline{\phi}|$ and Eq.~(\ref{VF1}) takes the form
\begin{equation}
V_F = \kappa^2 M^4 \left( 1 - \kappa_S \frac{|s|^2}{m_P^2} + \frac{1}{2} \gamma_S \frac{|s|^4}{m_P^4} + \cdots \right) + \kappa^2 |\phi|^2 \left( 2 \left( |s|^2 - M^2 \right) + \cdots \right) + \cdots,
\nonumber
\label{VF2}
\end{equation}
where 
\begin{equation}
\gamma_S \equiv 1 - \frac{7}{2}\kappa_S + 2\kappa_S^2 - 3\kappa_{SS}.
\label{gammaS}
\end{equation}
For simplicity, we will take $\kappa_S, \kappa_{SS} > 0$ throughout our calculations.  Suitable initial conditions ensure that the system of fields evolves along the valley of local minima located at $|\phi| = |\overline{\phi}| = 0$, $|s| > s_c = M$.  (For a detailed analysis of initial conditions in hybrid inflation, see Ref.~\citep{Clesse:2008pf}.)  As long as the couplings remain perturbative and $|s| < m_P$ (such that SUGRA corrections remain under control), the potential is dominated by the constant term $V_0 = \kappa^2 M^4$.  Thus $V>0$ and SUSY is broken during inflation, and radiative corrections and soft SUSY-breaking contributions must be taken into account.  It has previously been shown in the literature that each of these additional contributions can play an important role in the predictions of the model~\citep{Dvali:1994ms,Senoguz:2003zw,Senoguz:2004vu,Jeannerot:2005mc,urRehman:2006hu}.

It is convenient to reparametrize the field in terms of a dimensionless quantity, $x \equiv |s|/M$.  After including the radiative and soft corrections, and retaining terms up to order $|s|^4$ from the SUGRA contribution, the scalar potential during inflation becomes
%\begin{equation}
\begin{multline}
V \simeq \kappa^{2}M^{4} \left( 1 - \kappa_S \left( \frac{M}{m_P} \right)^2 x^2 + \gamma_S\left( \frac{M}{m_{P}}\right)^{4}\frac{x^{4}}{2} + \frac{\kappa ^{2}\mathcal{N}}{8\pi ^{2}}F(x) \right. \\
 \left. + a\left(\frac{m_{3/2}\,x}{\kappa\,M}\right) + \left( \frac{M_S\,x}{\kappa\,M}\right)^2\right) ,
\label{scalarpot}
\end{multline}
%\end{equation}
where 
\begin{equation}
F(x)=\frac{1}{4}\left( \left( x^{4}+1\right) \ln \frac{\left( x^{4}-1\right)}{x^{4}}+2x^{2}\ln \frac{x^{2}+1}{x^{2}-1}+2\ln \frac{\kappa ^{2}M^{2}x^{2}}{Q^{2}}-3\right)
\end{equation}
encompasses the radiative corrections~\citep{Dvali:1994ms}, and 
\begin{equation}
a = 2\left| 2-A\right| \cos [\arg s+\arg (2-A)]
\label{a}
\end{equation}
is the effective coefficient of the linear soft term having a coupling $(A-2)$ in the Lagrangian.  The coefficient $\mathcal{N}$ is the dimensionality of the gauge multiplets $\phi, \overline{\phi}$ under $G$, $Q$ is the renormalization scale, and we take a gravitino mass of $m_{3/2} \approx 1$~TeV.  The form of the soft SUSY-breaking terms (i.e. the last two terms appearing in Eq.~(\ref{scalarpot})) is derived from a gravity-mediated SUSY-breaking scheme, and both of the coefficients $a$ and $M_S^2$ may be either positive or negative.  Ref.~\citep{Rehman:2009yj} has shown that a negative soft mass-squared for the inflaton having a magnitude at intermediate scales can readily lead to good agreement with the WMAP central value of $n_s$.  For large tensor modes, we will see that an intermediate-scale soft mass remains important, yet in the present case we will have $M_S^2 > 0$ as in split supersymmetry models~\citep{ArkaniHamed:2004fb,*Giudice:2004tc,*ArkaniHamed:2004yi}.  We take $a$ at constant values, which can be achieved via an appropriate choice of initial conditions~\citep{urRehman:2006hu}.  The values we consider will be discrete, with the intention of representing the possible sign choices for $a$.

The vast majority of models in the literature treat the inflationary dynamics using the slow-roll approximation, in which cosmological quantities may be expanded in powers of the slow-roll parameters
\begin{eqnarray}
\epsilon &=& \frac{m_{P}^{2}}{4\,M^2}\left( \frac{V'}{V}\right) ^{2}, \label{epsilon} \\
\eta &=& \frac{m_{P}^{2}}{2\,M^2}\left( \frac{V''}{V}\right), \label{eta} \\
\xi^2 &=& \frac{m_P^4}{4M^4} \left( \frac{V' V'''}{V^2} \right) \label{xi} ,
%\label{slowroll}
\end{eqnarray}
where primes denote a derivative with respect to $x$.  Within the slow-roll approximation (i.e. for $\epsilon$, $|\eta|$, $\xi^2 \ll 1$), inflation lasts for a number of e-foldings given by
\begin{equation}
N_{0}=2\left( \frac{M}{m_{P}}\right) ^{2}\int_{x_e}^{x_{0}}\left( \frac{V}{V'}\right) dx ,
\label{N0}
\end{equation}
where $x_e$ parametrizes the field value at the end of inflation, and a subscript `0' corresponds to the pivot scale $k_0 = 0.002 \text{ Mpc}^{-1}$ crossing the horizon.  The value of $x_e$ is fixed either by the breakdown of the slow roll approximation, or by a `waterfall' destabilization occurring at the value $x_c = 1$ if the slow roll approximation holds.  To leading order, the scalar spectral index, tensor-to-scalar ratio, spectral running, and primordial curvature perturbation may (respectively) be written as
\begin{eqnarray}
n_s &\simeq& 1 - 6\epsilon + 2\eta, \label{nsSR} \\
r &\simeq& 16\epsilon, \label{rSR} \\
\frac{dn_s}{d\ln k} &\simeq& 16\epsilon\eta - 24\epsilon^2 - 2\xi^2, \label{alphaSR} \\
\Delta_{\mathcal{R}}^2 &\simeq& \frac{M^2}{6 \pi^2 m_{P}^{6}}\left( \frac{V^{3}}{(V')^2}\right). \label{curvSR}
\end{eqnarray}
For comparison with WMAP 7-year measurements~\citep{Komatsu:2010fb}, we will evaluate these functions at the pivot scale $x_0$.

In order to construct a viable model it is, of course, necessary to ensure that Eqs.~(\ref{N0})--(\ref{curvSR}) yield values in good keeping with the most recent experimental results.  However, if all of the independent parameters in the potential~(\ref{scalarpot}) are allowed to vary simultaneously, an analytical solution quickly becomes intractable.  (This is particularly true if we wish for some of the derived quantities to fall within a {\it range} of good values, rather than taking on the central value alone.)  One possibility for circumventing this issue is to make various simplifying assumptions to relate or eliminate some of the parameters.  It is, however, difficult to see how this approach may lead to a tensor-to-scalar ratio that is multiple orders of magnitude larger than what has been predicted in other treatments of this model.  In the next section, we will outline the approach we have employed to search for large $r$-values.

%%%%%%%%%%%%%%%%%%%%%%%%%%%%%%%%%%%%%%%%%%%%%%%%%

\section*{Seeking Large Tensor Modes}

Our goal is to determine whether there exists a region of the parameter space $\lbrace \kappa, M, \kappa_S, \kappa_{SS}, a, M_S, x_0 \rbrace$ specifying the potential in Eq.~(\ref{scalarpot}), which can lead to large primordial gravity waves.  For this, it is necessary to explore large regions of this many-dimensional space, and it is advantageous to place only the most conservative constraints in order to allow for a thorough investigation.  While sophisticated computational techniques would likely provide a great deal of useful insight, we have determined that a simpler approach would suit our criteria quite well.

In order to explore the parameter space, we have employed a `brute force' random generation of points in this space.  For each of the parameters in the potential, we have chosen a range within which to generate values randomly.  For parameters which were expected to vary over multiple orders of magnitude, we have instead chosen to randomly generate the base-10 logarithm of the parameter (e.g. if we allow $\kappa$-values on the interval $[10^{-4},1]$, it is perhaps more useful to generate $\log\kappa$ on the interval $[-4,0]$).  Table~\ref{ranges} shows the ranges of the fundamental and derived quantities that will correspond to the figures in this letter.

\begin{table}
\begin{tabular}{c|c|c||c|c}
%$\begin{array}
{\bf Fundamental} & {\bf Range} & {\bf Scale} & {\bf Derived} & {\bf Constraining range} \\
{\bf parameter} &  & {\bf type} & {\bf quantity} &  \\
\hline
$\kappa$ & $[10^{-4}, 5]$ & log & $n_s$ & $[0.920, 1.016]$ \\
$M/m_P$ & $[10^{-4}, 10^{-1}]$ & log &  & $ = 0.968 \pm 4\sigma$ \\      %$\Delta_{\mathcal{R}}^2$ & $[2.21, 2.65] \times 10^{-9} = (2.43 \times 10^{-9} \pm 2\sigma)$ \\
\cline{4-5}
$M_S/m_P$ & $[10^{-8}, 10^{-4}]$ & log & $\Delta_{\mathcal{R}}^2$ & $[2.21, 2.65] \times 10^{-9}$ \\      %$r$ & $< 1$ \\
$\kappa_S$ & $[10^{-5}, 3]$ & log &  & $= 2.43 \times 10^{-9} \pm 2\sigma$ \\      %$N_0$ & $[50,60]$ \\
\cline{4-5}
$\kappa_{SS}$ & $[10^{-5}, 3]$ & log & $r$ & $< 1$ \\
$a$ & $\lbrace -1,0,1 \rbrace$ & --- & $N_0$ & $[50,60]$ \\
$x_0$ & $[1, \frac{m_P}{M}]$ & linear &  &  \\

%\end{array}$
\end{tabular}
\caption{Ranges specified for the fundamental parameters in Eq.~(\ref{scalarpot}), and constraints placed on derived quantities.  Note that $a$ was considered at discrete values, and $x_0$ can take on any value between the waterfall point and the Planck scale.  Central values and standard deviations for measured quantities are in reference to the WMAP 7-year analysis~\citep{Komatsu:2010fb}.}
\label{ranges}
\end{table}

Each set of values $\lbrace \kappa, M, \kappa_S, \kappa_{SS}, a, M_S, x_0 \rbrace$ represents a point in parameter space, and specifies a unique potential via Eq.~(\ref{scalarpot}).  After generating these values for a given point, functions derived from the potential can be calculated with ease.  Before considering whether the point fits our desired constraints, we check a handful of basic properties.  For instance, the physical inflaton mass-squared
\begin{equation}
m_{\text{inf}}^2 = \kappa^2 M^2 \left[ 2 - \kappa_S \left( \frac{M}{m_P} \right)^2 \right] + M_S^2
\label{minfsq}
\end{equation}
is required to be positive at the global minimum of the potential, $(s, \phi = \overline{\phi}) \rightarrow (0, M)$.  (While such considerations are built into the physics of the model, there is no guarantee that the points randomly generated in parameter space will lead to physical results, and so this type of constraint must be imposed by hand.)  Also, since we expect the region of interest to coincide with large values of the field, the linear term will be suppressed unless $a$ is very large.  Thus while we retain this term in our calculations, it can effectively be ignored for qualitative (and most quantitative) considerations.

In addition, it is necessary to place constraints in order to ensure that the fields evolve to the (SUSY) global minimum.  In particular, whenever negative contributions to the potential are present, there exists a possibility of a false (metastable) vacuum developing.  If the inflaton is trapped in such a vacuum state, inflation may last for a very long time and SUSY will remain broken.  In order to ensure that the inflaton evolves to the SUSY minimum, we require that the potential be essentially monotonic over the interval $x \in [1, x_0]$.

An immediate consequence of the assumption that the potential is monotonic over a given range is a limitation on the behavior of the polynomial terms in $V$.  (The term arising from radiative corrections yields only a subtle contribution to $V'$, especially at large values of $x$.)  To facilitate this discussion, we may rewrite the potential in the form
\begin{equation}
\frac{V}{V_0} \supset \left( -\kappa_S + \frac{M_S^2 m_P^2}{V_0} \right) y^2 + \frac{\gamma_S}{2} y^4,
\label{Vpoly}
\end{equation}
where we have defined $y \equiv |s|/m_P$, with $y \leq 1$ in order that SUGRA corrections remain under control.  We expect large $r$-values to be obtained for $y_0 \simeq 1$, where the quadratic and quartic terms may come into competition depending on the size of their coefficients; the quadratic term dominates unless $|\gamma_S|$ is rather large.  Eq.~(\ref{gammaS}) tells us that $\gamma_S$ may in principle be positive or negative.  For $\kappa_S, \kappa_{SS}$ of order unity or smaller, a large positive $\gamma_S$ can only be obtained for small values of both couplings; then small values of $M_S^2$ are required to obtain an overall quadratic coefficient that is negative.  For $\gamma_S < 0$, the quadratic coefficient must be positive, and the combination $\frac{7}{2}\kappa_S + 3\kappa_{SS}$ cannot be too large lest the quartic term begin to dominate for $x < x_0$.  (A similar potential having positive quadratic and negative quartic contributions was treated in Ref.~\citep{Lin:2008ys}.)  The case where both the quadratic and quartic terms are positive typically leads to $n_s > 1$, owing to $\eta > 0$ in Eq.~(\ref{nsSR}).

Continuing in this same vein, it will be informative to derive an approximate analytical expression for $r$.  As we have already argued, for large $y_0$ we may suppress the radiative correction and soft linear terms.  Using Eq.~(\ref{Vpoly}), we may write
\begin{eqnarray*}
\frac{V'}{V_0} &\simeq& 2 \left(\frac{M}{m_P}\right) \cdot y \left[ \left( -\kappa_S + \frac{M_S^2 m_P^2}{V_0} \right) + \gamma_S y^2 \right] , \\
\frac{V''}{V_0} &\simeq& 2 \left(\frac{M}{m_P}\right)^2 \left[ \left( -\kappa_S + \frac{M_S^2 m_P^2}{V_0} \right) + 3\gamma_S y^2 \right] .
\end{eqnarray*}
Approximating $V(x_0) \simeq V_0$ and using Eq.~(\ref{eta}) allows us to eliminate the quadratic coefficient in favor of $\eta$:
\begin{equation}
\frac{V'}{V_0} \simeq 2 \left(\frac{M}{m_P}\right) \cdot y \left[ -2 \gamma_S y^2 + \eta \right].
\label{Vpapprox}
\end{equation}
It is worth pausing at this intermediate step to draw a few conclusions.  We see that, in the large-$r$ regime, the sign of $V'$ is determined by a competition of two terms.  (It turns out that, despite its required smallness during inflation, $\eta$ is of the same order as the other term here and thus cannot be omitted.)  Then the requirement that the potential be monotonic up until $x = x_0$ leads to
\begin{equation*}
-2 \gamma_S y_0^2 + \eta \geq 0 .
\end{equation*}
For $\eta > 0$, $\gamma_S$ can be positive, but the largeness of $y_0$ and necessity of $|\eta| \ll 1$ demand that $|\gamma_S|$ be very small.  On the other hand, $\eta < 0$ requires $\gamma_S < 0$, and $|\gamma_S|$ is not limited by the size of $|\eta|$.  That being said, if $|\gamma_S|$ is too large a local maximum may develop for $x$ significantly removed from the Planck scale.  Since we require a monotonic potential for $x < x_0$, this drives $x_0$ to smaller values and radiative corrections can no longer be reliably suppressed.  More stringently, smaller $y_0$ is expected to lead to smaller values of $r$.  Thus we expect large $r$-values to arise primarily for $\gamma_S$ negative and of modest magnitude.

\begin{figure}[tb]
\centering
\includegraphics[width=10cm]{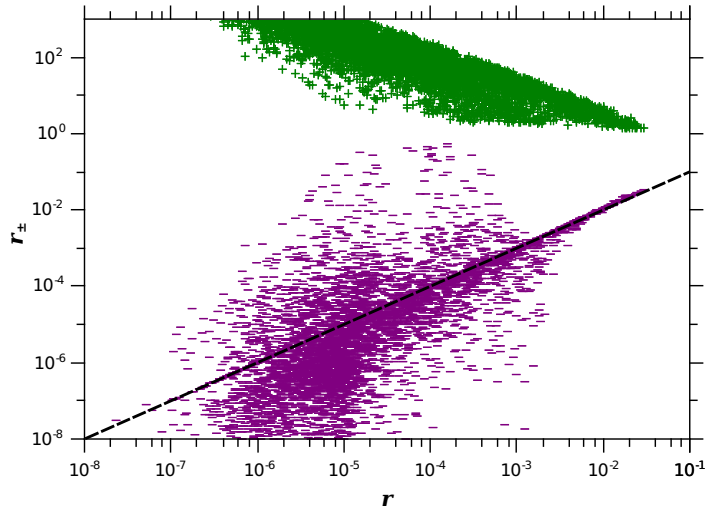}
\caption{Approximate values $r_\pm$ (see Eq.~(\ref{rapprox_pm})) vs. calculated values $r$ of the tensor-to-scalar ratio.  The two solutions $r_+$ and $r_-$ are represented by green crosses and purple horizontal lines, respectively, with a black dashed line denoting equality with $r$.  We see that $r_-$ is the appropriate choice, yielding an excellent approximation for large tensor modes.  The results displayed here correspond to $G = \text{U(1)}$, but the conclusion holds equally well in the flipped SU(5) case. }
\label{rapprox_fig}
\end{figure}

With these considerations in mind, we may write down an approximate formula for $r$ using Eqs.~(\ref{epsilon},~\ref{rSR},~\ref{Vpapprox}):
\begin{equation}
r \simeq 16 y_0^2 \left[ -2 \gamma_S y_0^2 + \eta \right]^2 .
\label{rapprox_eta}
\end{equation}
The value of $\eta$ at the beginning of inflation is determined in a complicated way by the variation of many parameters.  It may be useful to seek a substitution for a parameter whose value is better known.  Indeed, using Eqs.~(\ref{nsSR},~\ref{rSR},~\ref{rapprox_eta}), we find
\begin{equation}
r_\pm \simeq \frac{8}{9 y_0^2} \left[ 1 + 3y_0^2 (1-n_s) + 12\gamma_S y_0^4  \pm  \sqrt{1 + 6y_0^2 (1-n_s) + 24\gamma_S y_0^4} \right] .
\label{rapprox_pm}
\end{equation}
It turns out that, after satisfying the full set of constraints, $r_+$ is quite large; indeed, in most cases this solution violates the slow roll approximation that we have assumed in this derivation, as can be seen in Fig.~\ref{rapprox_fig}.  In contrast, $r_-$ retains its validity and is an excellent fit in the region where $r$ is large.

%%%%%     FIGURE

%%%%%%%%%%%%%%%%%%%%%%%%%%%%%%%%%%%%%%%%%%%%%%%%%

\section*{Results and Discussion}

%%%%%     U(1)

In order to perform the calculations, it is necessary to make a choice of gauge group $G$, which determines the size of the radiative correction term in the potential.  We will chiefly discuss the simplest choice of $G$, namely U(1) (i.e. $\mathcal{N} = 1$), which may be identified with a $B - L$ gauge symmetry.  Some degree of caution must be employed, however, due to the formation of cosmic strings upon the breaking of $G$ at the end of inflation.  Indeed, since $G\mu \sim (M/m_P)^2$~\citep{Vilenkin_book} and since large $r$ implies large vacuum energy $V_0^{1/4} = \sqrt{\kappa} M$, we immediately expect high string tension (larger than the current bound $G\mu \lesssim (2$--$7) \times 10^{-7}$~\citep{Battye:2006pk,*Battye:2010xz}) to accompany large tensor modes.  To address this, one may employ a modified framework such as the so-called shifted hybrid inflation~\citep{Jeannerot:2000sv,Senoguz:2003zw}, for the purpose of inflating away the strings.  Although we do not pursue this option, we expect the shifted inflation case to yield qualitatively similar results to those presented here.  An alternative is to make a different choice for $G$, such as flipped SU(5)~\citep{Li:2010rz}, which we will discuss at the end of this section.

\begin{figure}[pt]
\centering
\begin{tabular}{cc}
\multicolumn{2}{c}{Results for $G = \text{U(1)}$} \\
\includegraphics[width=6cm]{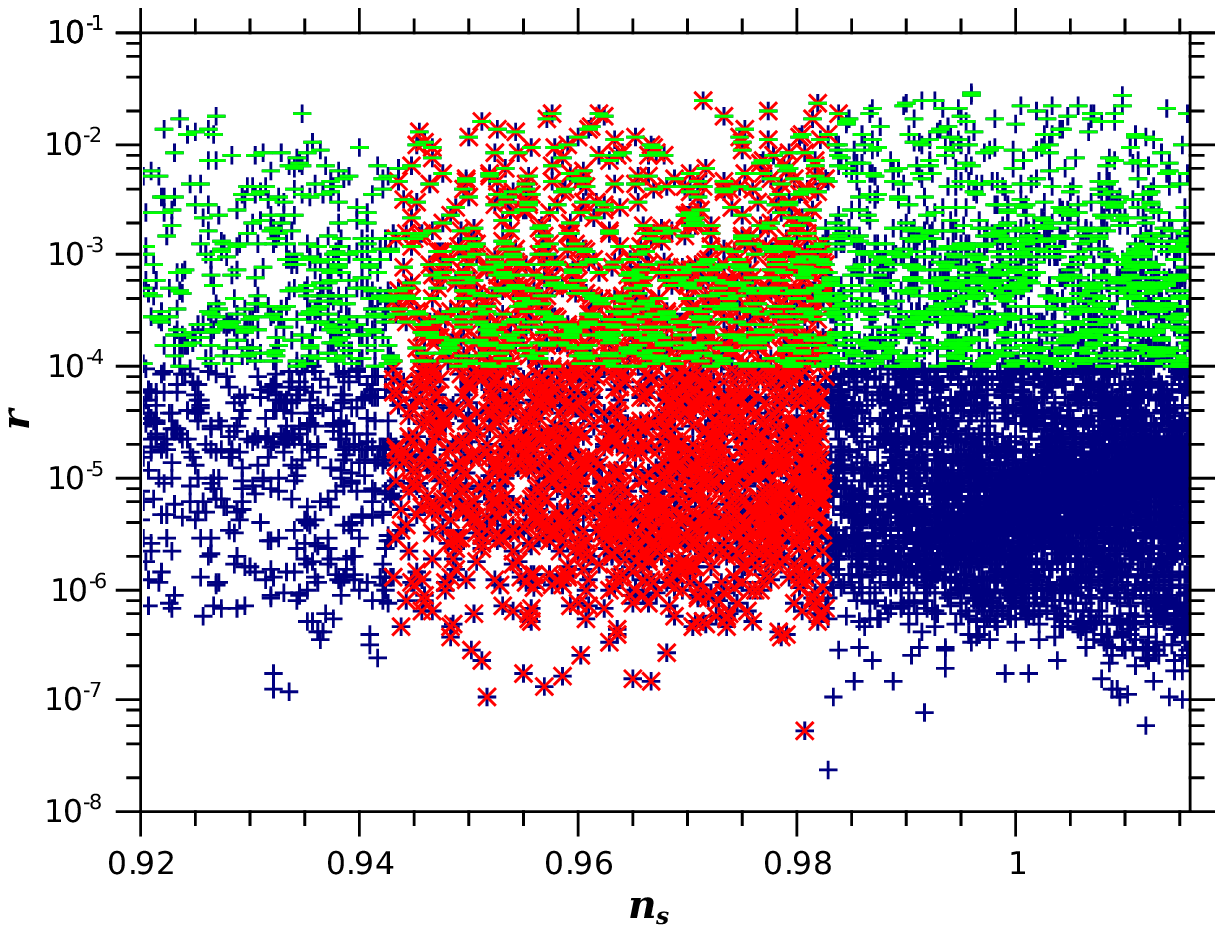} &
\includegraphics[width=6cm]{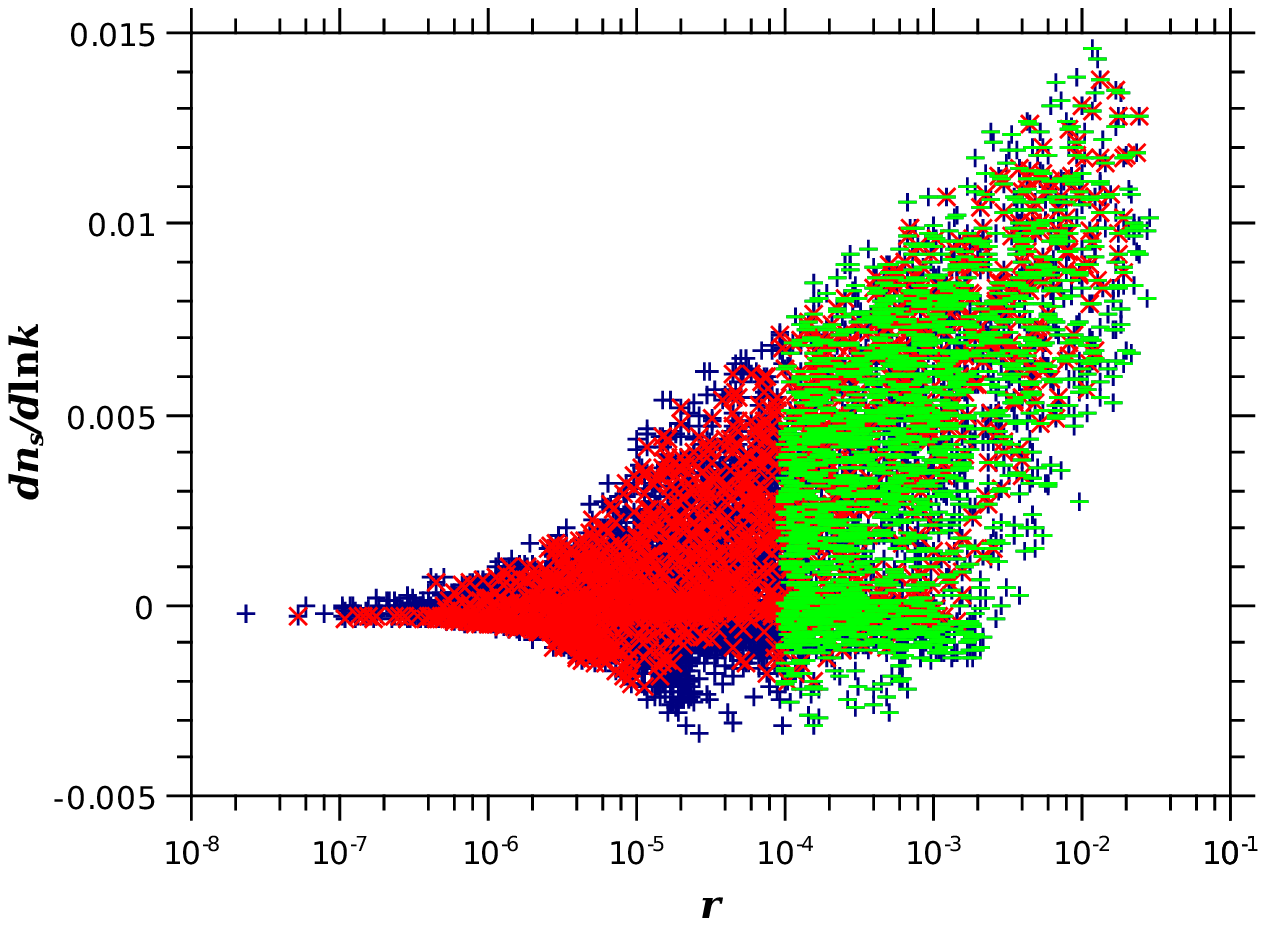} \\
(a) & (b) \\
\includegraphics[width=6cm]{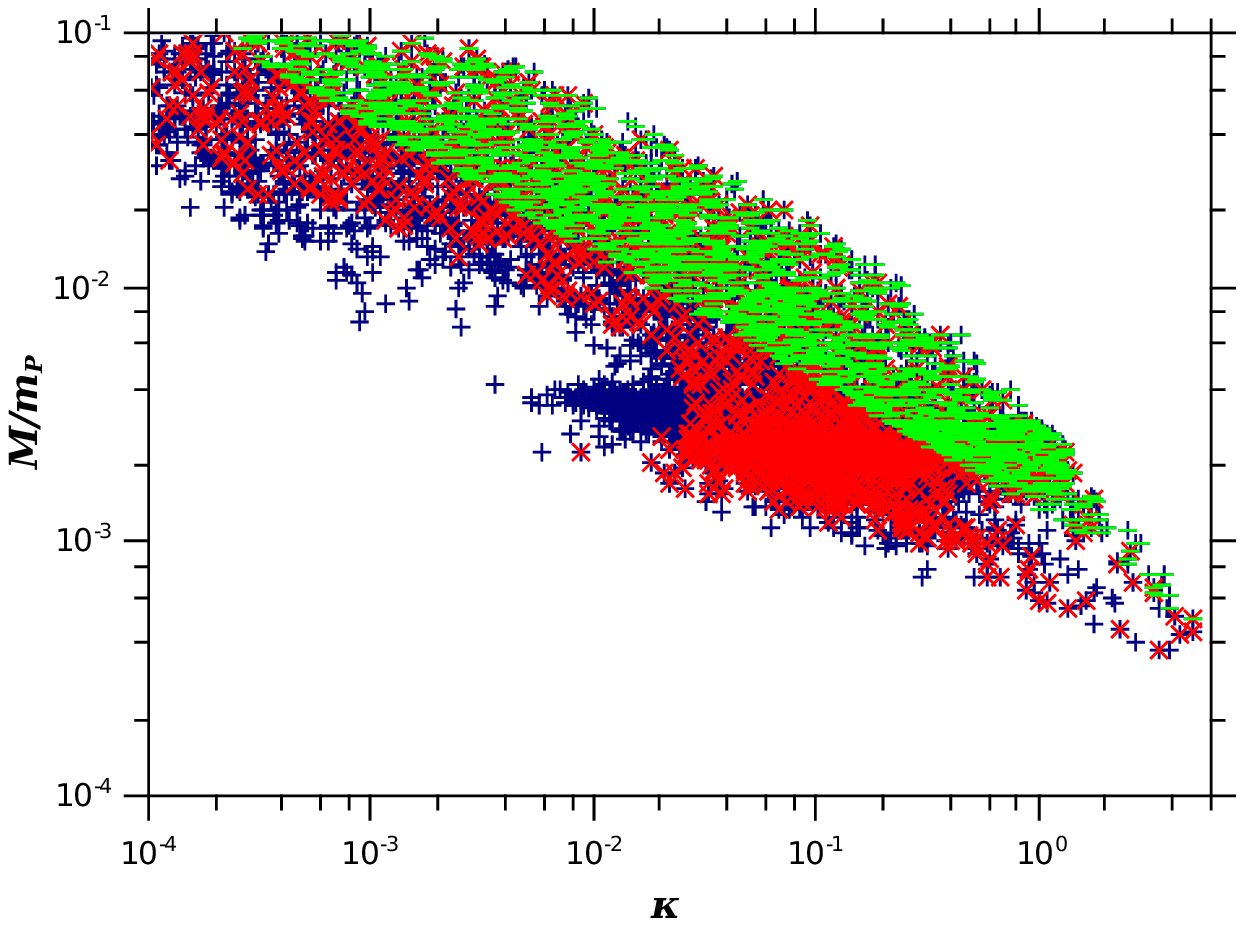} &
\includegraphics[width=6cm]{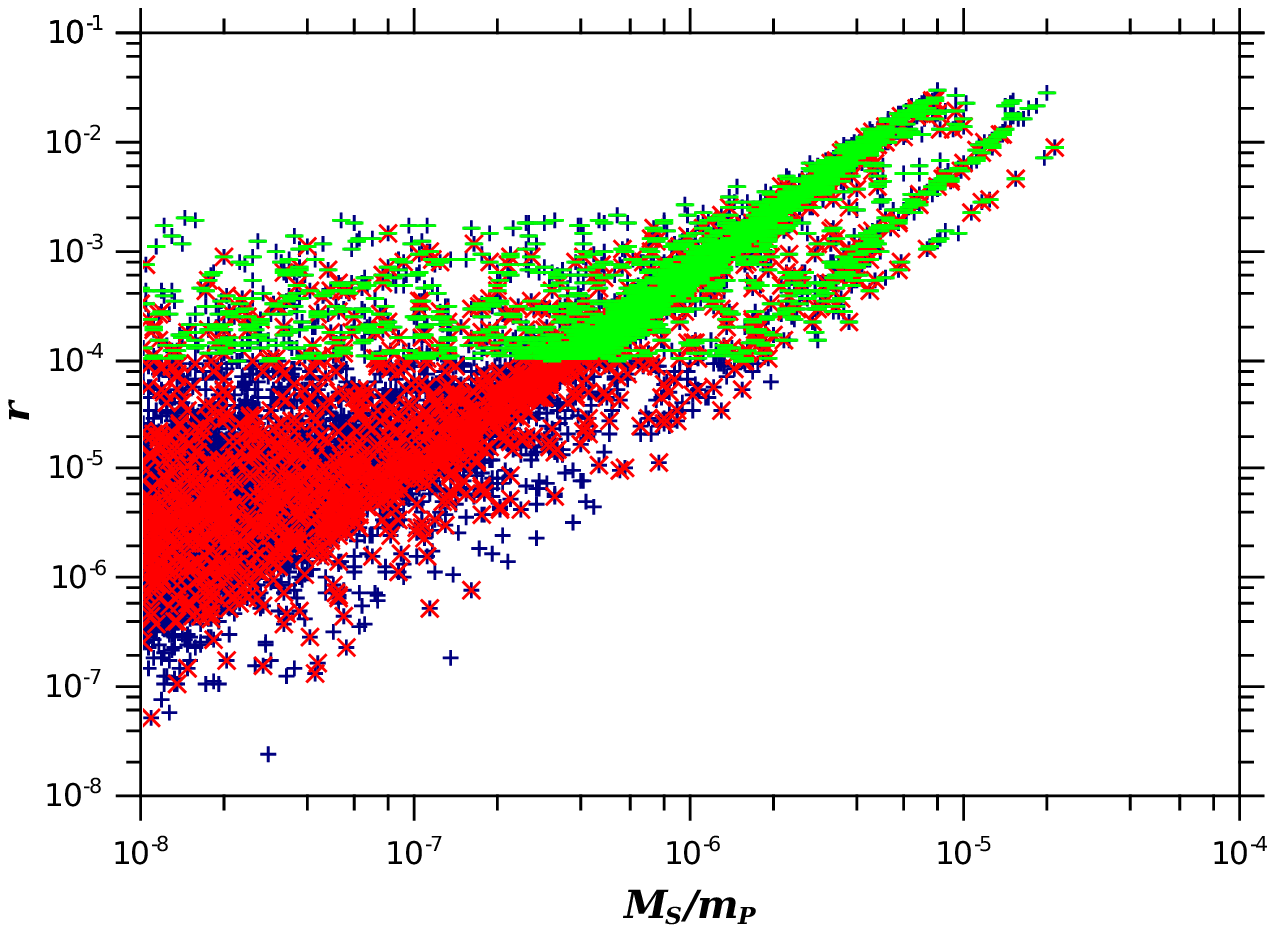} \\
(c) & (d) \\
\includegraphics[width=6cm]{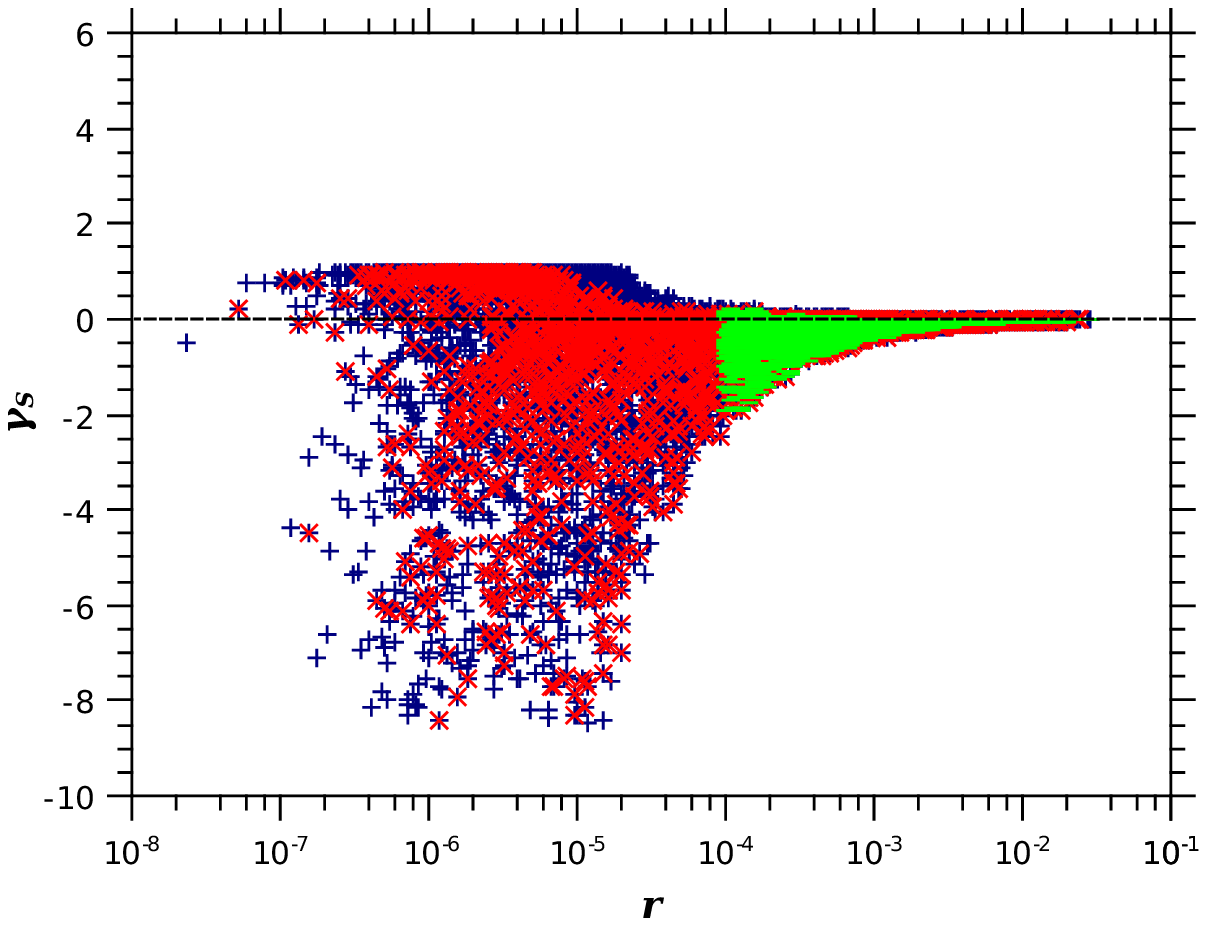} &
\includegraphics[width=6cm]{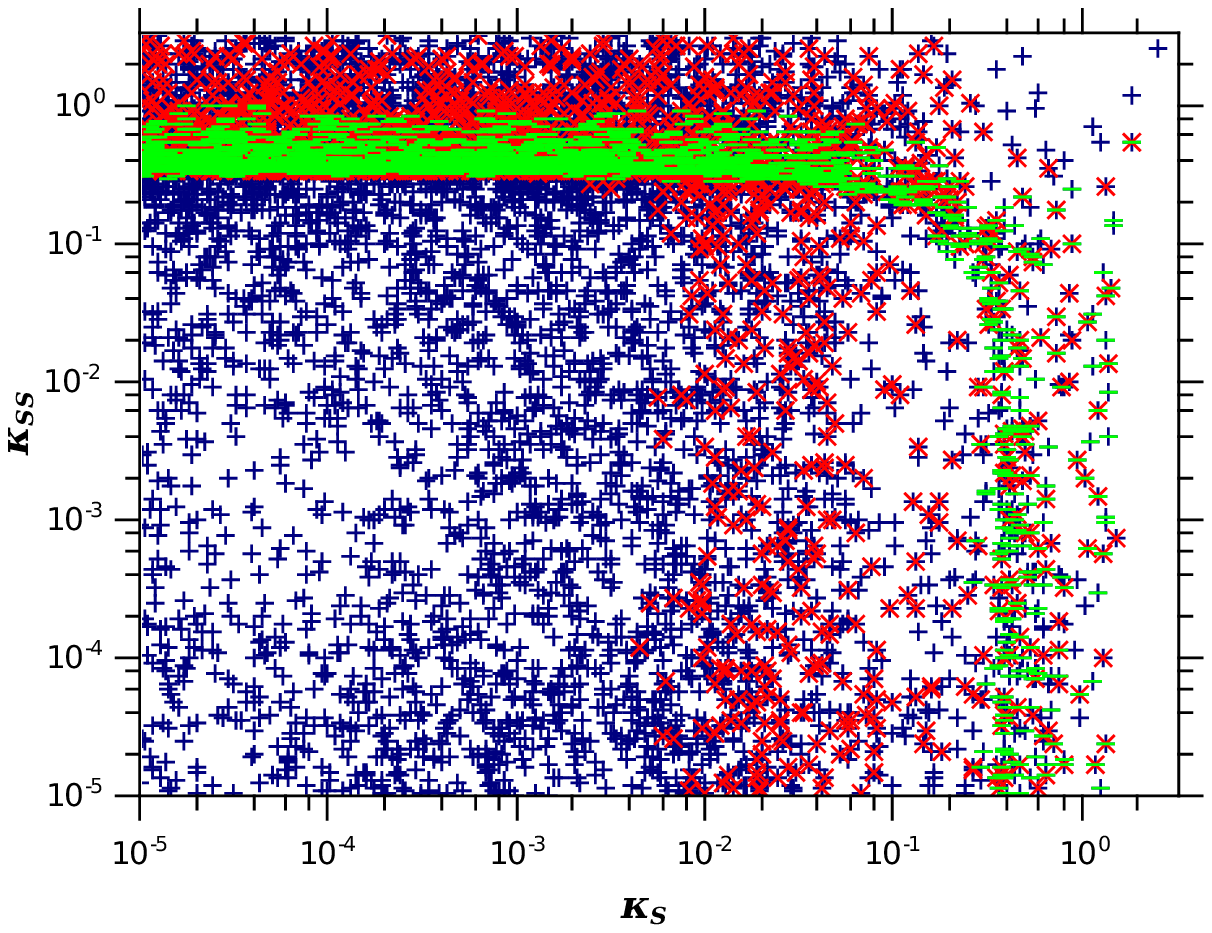} \\
(e) & (f) \\
\end{tabular}
\caption{Results of the numerical calculations for the U(1) case.  As demonstrated in panel (a), the points are color- and symbol-coded to represent regions of interest.  Blue crosses represent all points generated according to Table~\ref{ranges}; red X's correspond to a 1$\sigma$ range of $n_s$; and green horizontal lines signify points with `large' $r$-values ($\gtrsim 10^{-4}$).}
\label{figpanel_NN1}
\end{figure}

The results of our numerical calculations using $G = \text{U(1)}$ are presented in Fig.~\ref{figpanel_NN1}.  Panel~(a) displays the behavior of the tensor-to-scalar ratio with respect to the spectral index.  We see that these parameters are essentially uncorrelated in our model; in particular, large values of $r$ can be obtained for essentially any $n_s$ value within the 4$\sigma$ range we have explored.  It is especially worth pointing out that this model has no difficulty in generating a spectral index near the WMAP7 central value $n_s \simeq 0.968$.  This is due largely to negative values of $V''$ at the start of inflation.  Referring to Eq.~(\ref{nsSR}) and noting that (typically) $\epsilon \ll |\eta|$, we see that $\eta < 0$ will drive the spectral index down to red-tilted values ($n_s < 1$).  In this regard, these SUSY hybrid models resemble models of hilltop inflation~\cite{Boubekeur:2005zm,*Kohri:2007gq,*Pallis:2009pq,Lin:2008ys}.

%%%%%     FIGURE PAGE for U(1)

The first panel of Fig.~\ref{figpanel_NN1} also serves to define the color- and symbol-coding used in the other panels.  It is useful, for example, to see that points having $n_s$ within 1$\sigma$ of the central value (red X's) cover essentially the same regions as the 4$\sigma$ points (blue crosses) in many slices of the parameter space.\footnote{As can be seen in Table~\ref{ranges}, the 4$\sigma$ range was imposed by taking (central value $\pm 4\sigma$), as an initial cut to restrict the results to interesting values.  In contrast, what we refer to here as the 1$\sigma$ range corresponds to the region bounded by the 1$\sigma$ curves in the $(n_s, r)$ plane as presented in Ref.~\citep{Komatsu:2010fb}.  These points were extracted upon analysis of the 4$\sigma$ results, at which stage a reduced number of points made it realistic to use a more precise approach.}  Perhaps even more telling is the location of the points with large $r$-values, denoted by green horizontal lines and for which we have chosen $r \gtrsim 10^{-4}$ as an arbitrary cutoff.

In panel~(b), we show the dependence of the spectral running $dn_s/d\ln k$ on $r$.  While the largest $r$-values can lead to an alarmingly large running, this is not necessarily the case; it is also possible to generate large tensor modes for much smaller magnitudes of the spectral running.  It is not {\it a priori} obvious which case is more likely.  As our aim is to explore the parameter space, we simply note that there exists a possibility for observable $r$ with $dn_s/d\ln k$ in keeping with current experimental results.

Panel~(c) shows an overall inverse correlation between $\kappa$ and $M$, which is expected from the estimate $V_0^{1/4} \equiv \sqrt{\kappa} M \sim 10^{15 \text{--} 16}$~GeV.  The well-known slow-roll formula $V(x_0)^{1/4} \simeq 3.3 \times 10^{16} \text{ GeV} \cdot r^{1/4}$, with $V_0 \sim V(x_0)$, dictates that $r$ should obtain its maximal values when one of $\kappa$, $M$ is large.  Based upon the reasoning put forth in Refs.~\citep{urRehman:2006hu,Rehman:2009nq,Rehman:2009yj}, one may anticipate large tensor modes to occur for large values of $\kappa$.  However, we see here that large $r$ is obtained quite readily for moderate values of $\kappa$, corresponding instead to $M$-values larger than anticipated, and even approaching the Planck scale.

From panel~(d), it appears that $r$ and $M_S$ are somewhat correlated, especially for $r \gtrsim 10^{-3}$.  This can be understood by noting that the quadratic coefficient must be large enough to overcome the influence of the (negative) quartic term and stabilize the potential, as discussed in the previous section.  In order for this to be true as $y_0$ (and thus $r$) increases, $M_S$ must also increase substantially, and becomes quite sizable as the field amplitude approaches the Planck scale.  In these models, we find $M_S/m_P \lesssim 10^{-5}$ (or $M_S \lesssim 10^{13 \text{--} 14}$~GeV), much larger than the soft mass values utilized in many implementations of supersymmetric models.  Indeed, these intermediate mass scales are reminiscent of split-SUSY models~\citep{ArkaniHamed:2004fb}, in which the scalar soft masses may lie many orders of magnitude above the fermionic soft masses, which remain around TeV-scale.

Panels~(e) and (f) shed light on the behavior of parameters related to the non-minimal K\"ahler potential.  Panel~(e) supports our previous claim that $\gamma_S$ should be negative (and fairly small) at the largest values of $r$.  In panel~(f), we see that large $r$-values are mainly obtained for large $\kappa_{SS}$, in particular $\kappa_{SS} \gtrsim 1/3$ for which $\gamma_S < 0$ at small $\kappa_S$.  It is also interesting to point out that if both $\kappa_S$ and $\kappa_{SS}$ are sufficiently small, $n_s$ cannot lie within the 1$\sigma$ range and $r \gtrsim 10^{-4}$ is not possible.  In other words, the use of a non-minimal K\"ahler potential is critical for obtaining large tensor modes, particularly with favorable $n_s$.

As can be seen in Fig.~\ref{rapprox_fig} and panels~(a), (b), (d), (e) and (g) of Fig.~\ref{figpanel_NN1}, our calculations for $G = \text{U(1)}$ have yielded values of the tensor-to-scalar ratio up to $r \simeq 0.03$, which essentially coincides with the estimated threshold for detection by the current Planck satellite experiment.  If detected, tensor fluctuations in the CMB can give much information about the energy scale of inflation as well as gravitational wave amplitudes in the early universe.  A measurement by Planck lying close to this threshold could serve to support the validity of SUSY hybrid inflation models, yet a detection of $r$ significantly higher than 0.03 would be difficult to reconcile with the predictions of the model that we have presented here.

%%%%%     general considerations

The so-called `$\eta$-problem' has materialized as a stumbling block to many SUSY models of inflation, and as such, it is germane to discuss this issue in the context of the current model.  In SUGRA hybrid inflation with a minimal K\"ahler potential, this problem is eliminated by a fortuitous cancellation of the troublesome mass term with a term arising from the superpotential~\citep{Copeland:1994vg,Lazarides:2001zd}.  Generalizing to a non-minimal form of the K\"ahler potential introduces a new mass term for the inflaton, whose contribution cannot be suppressed for $\kappa_S$ non-negligible.  In our numerical calculations, we have required $\eta \leq 1$ for long enough to produce 50--60 e-foldings of inflation, i.e. the $\eta$-problem is not present in the numerical results we have presented in this letter.  This is accomplished by a delicate cancellation of terms in $V''(x_0)$, namely those arising from the quadratic and quartic terms in the potential.

It should also be noted that the behavior described in this letter is, at least in part, a consequence of our choice of signs for various parameters.  For example, our simplification $\kappa_S, \kappa_{SS} > 0$ leads to the restriction $M_S^2 > 0$ as well as affecting the behavior of the quartic coefficient $\gamma_S$.  Accordingly, it is possible that there exist other regions of parameter space leading to sizable $r$ which correspond to different sign arrangements of the couplings.  We have tested this in limited capacity, and while no such regions were revealed, a more thorough investigation is needed.  It would be particularly interesting if such sign changes could result in large tensor modes for $M_S \sim 1$~TeV.  We find that this may be possible for the choice $\kappa_S < 0$, $\kappa_{SS} > 0$, while other sign choices come into conflict with our previous assumptions.  In this case, however, it is unclear whether large $r$-values can result from perturbative values of the couplings.  Also, the drastic reduction in the magnitude of $M_S$ may lead to the soft linear term playing a more important role and thus altering the results.

The dynamics of the reheating phase can be an important consideration in any inflationary model.  This is especially true in models that include supergravity, where overproduction of TeV-scale gravitinos can spoil the success of Big Bang nucleosynthesis.  In our case, large $r$ can readily correspond to a large inflaton mass, which in turn can lead to a high gravitino density via thermal processes~(see Refs.~\citep{Lazarides:2001zd,Lyth:1998xn,*Mazumdar:2010sa} for reviews) and through non-thermal decays of the inflaton~\citep{Nakayama:2010xf}.  Possible mechanisms for suppressing the gravitino abundance include a period of thermal inflation and subsequent dilution by entropy production~\citep{Lazarides:1985ja}.  In this case, we may consider an additional symmetry whose breaking occurs spontaneously at an intermediate scale, while the corresponding phase transition is delayed to low energy scales.  Such a mechanism will also lead to suppression of the baryon asymmetry~\citep{Lazarides:1985ja}, such that the typical leptogenesis schemes~\citep{Fukugita:1986hr} (for non-thermal leptogenesis, see Ref.~\citep{Lazarides:1991wu,*Lazarides:1996dv}) will be insufficient; then, the lepton asymmetry should be over-generated in order to emerge at the appropriate abundance after dilution.  This may be the case, for example, in models of resonant leptogenesis~\citep{Pilaftsis:2003gt,*DeSimone:2007pa}.

While the Lyth bound given in Eq.~(\ref{lyth1}) applies to a generic inflation model, a more stringent version
\begin{equation}
r \lesssim 0.003 \left( \frac{50}{N_0} \right)^2 \left( \frac{\Delta\phi}{m_P} \right)^2,
\label{lyth2}
\end{equation}
may be used in a large class of models~\cite{Boubekeur:2005zm,Lyth:2009zz}.  The additional assumption here is that the relative slope $|V'/V|$, and hence also $\epsilon$, is monotonically increasing.  While this condition is true in many of the inflation models in the literature, it need not hold in general; indeed, one may readily verify that Eq.~(\ref{lyth2}) cannot yield values $r \sim 0.03$ for sub-Planckian values of the field.  Then the model currently being considered must violate this assumption in order to obtain the largest (and observationally most interesting) values of $r$ that we have generated.\footnote{ We thank David Lyth for drawing our attention to this feature of our model, and its possible implications including the following discussion on overproduction of primordial black holes.}  This essentially corresponds to a sign change of $V''$ along the inflationary trajectory, causing $\epsilon$ to decrease over some region of $x$.  (For a class of models exploiting a non-monotonic variation of $\epsilon$, see Ref.~\citep{BenDayan:2009kv}.)

In hybrid models, inflation may end via waterfall (induced by the dynamics of $\phi, \overline{\phi}$) and it is not necessary for $\epsilon$ to increase to unit order.  However, a drastic decrease in $\epsilon$ will cause $\Delta_{\mathcal{R}}^2 \sim V/\epsilon$ to increase dramatically, and one may begin to worry about the overproduction of primordial black holes (PBHs).  Conventional wisdom dictates that the amplitude of the curvature spectrum $\Delta_{\mathcal{R}}$ should not exceed unity at the end of inflation (when excess PBHs can no longer be inflated away).  We have checked our numerical results and verified that $\Delta_{\mathcal{R}} \lesssim 10^{-3}$, so that PBH overproduction does not concern us here.  This may be qualitatively understood by considering Eq.~(\ref{N0}) in the form $N_0 \sim \int_{x_e}^{x_{0}} dx/\sqrt{\epsilon}$.  If $\epsilon$ is very small over a significant range of $x$, we see that $N_0$ will tend to drift outside of the usual 50--60 window unless the range of integration $(x_0 - x_e)$ is narrowed.  But this is proportional to the $\Delta \phi$ contained in Eqs.~(\ref{lyth1}) and (\ref{lyth2}), and a marked decrease in this range will drive $r$ to smaller values as well, possibly into a region compatible with Eq.~(\ref{lyth2}) for which no region of decreasing $\epsilon$ is needed.  Incidentally, an overdensity of PBHs could also be diluted by entropy production in the same way as gravitinos, as described above.

%%%%%     FIGURE PAGE for flipped SU(5)

\begin{figure}[pt]
\centering
\begin{tabular}{cc}
\multicolumn{2}{c}{Results for $G = \text{SU(5)} \times \text{U(1)}_X$} \\
\includegraphics[width=6cm]{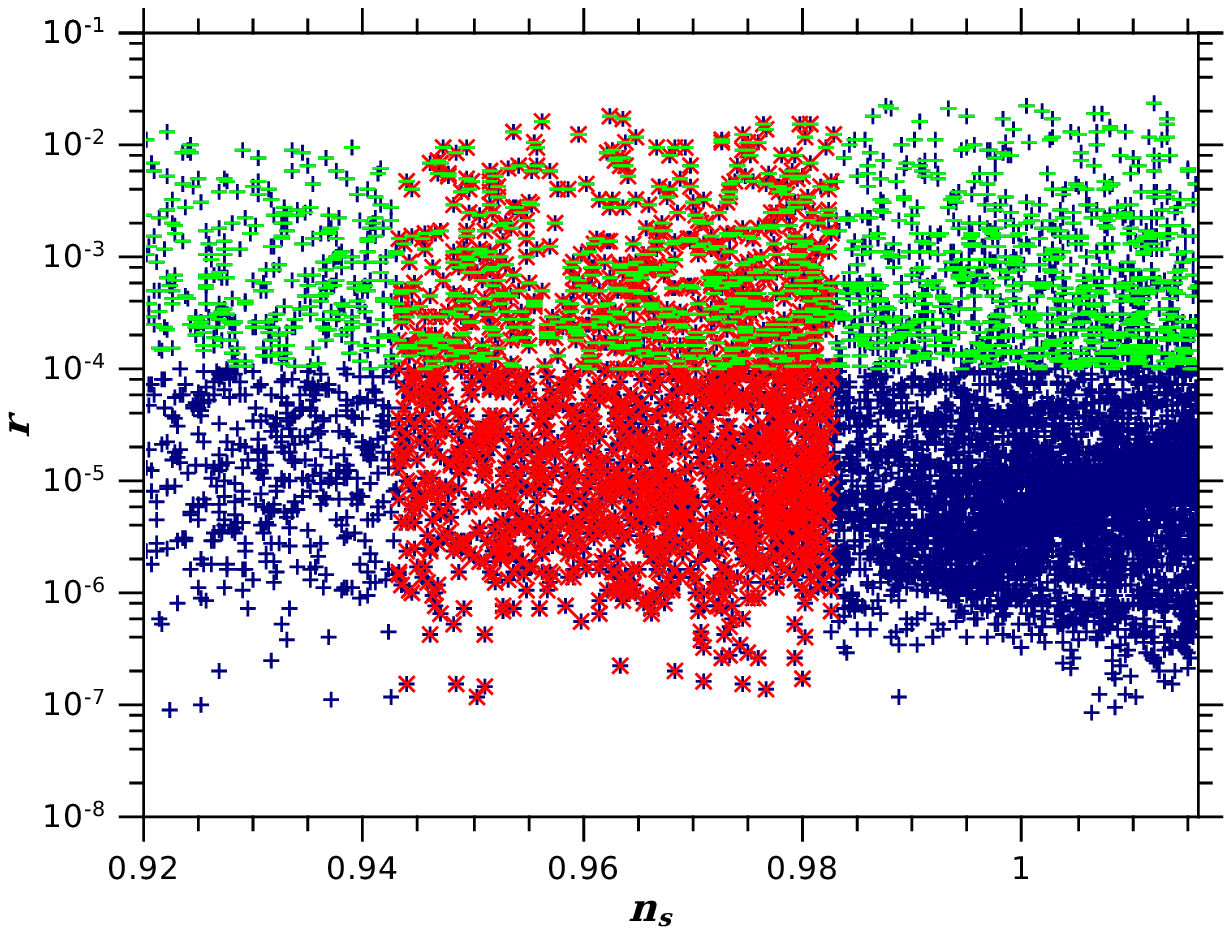} &
\includegraphics[width=6cm]{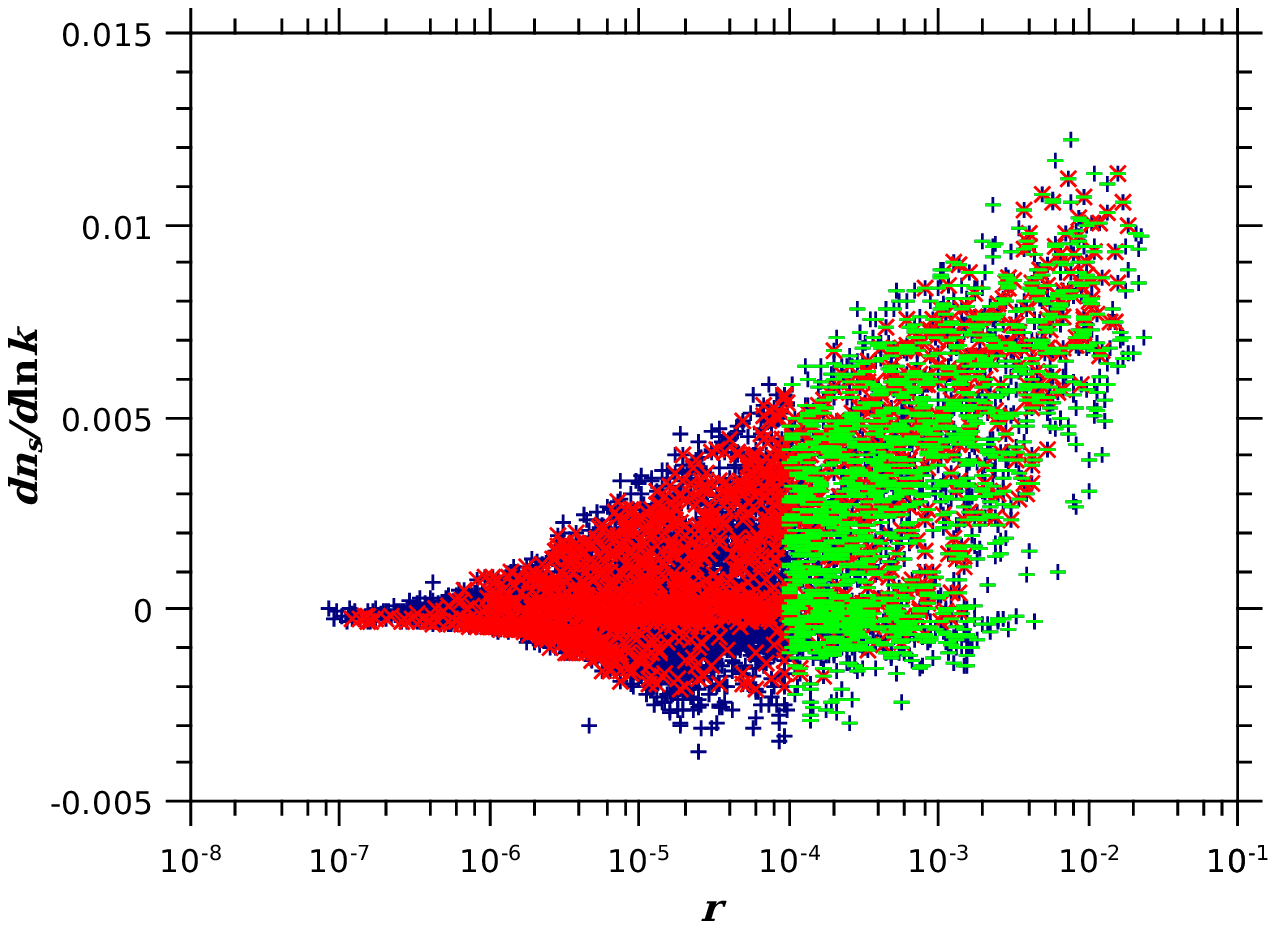} \\
(a) & (b) \\
\includegraphics[width=6cm]{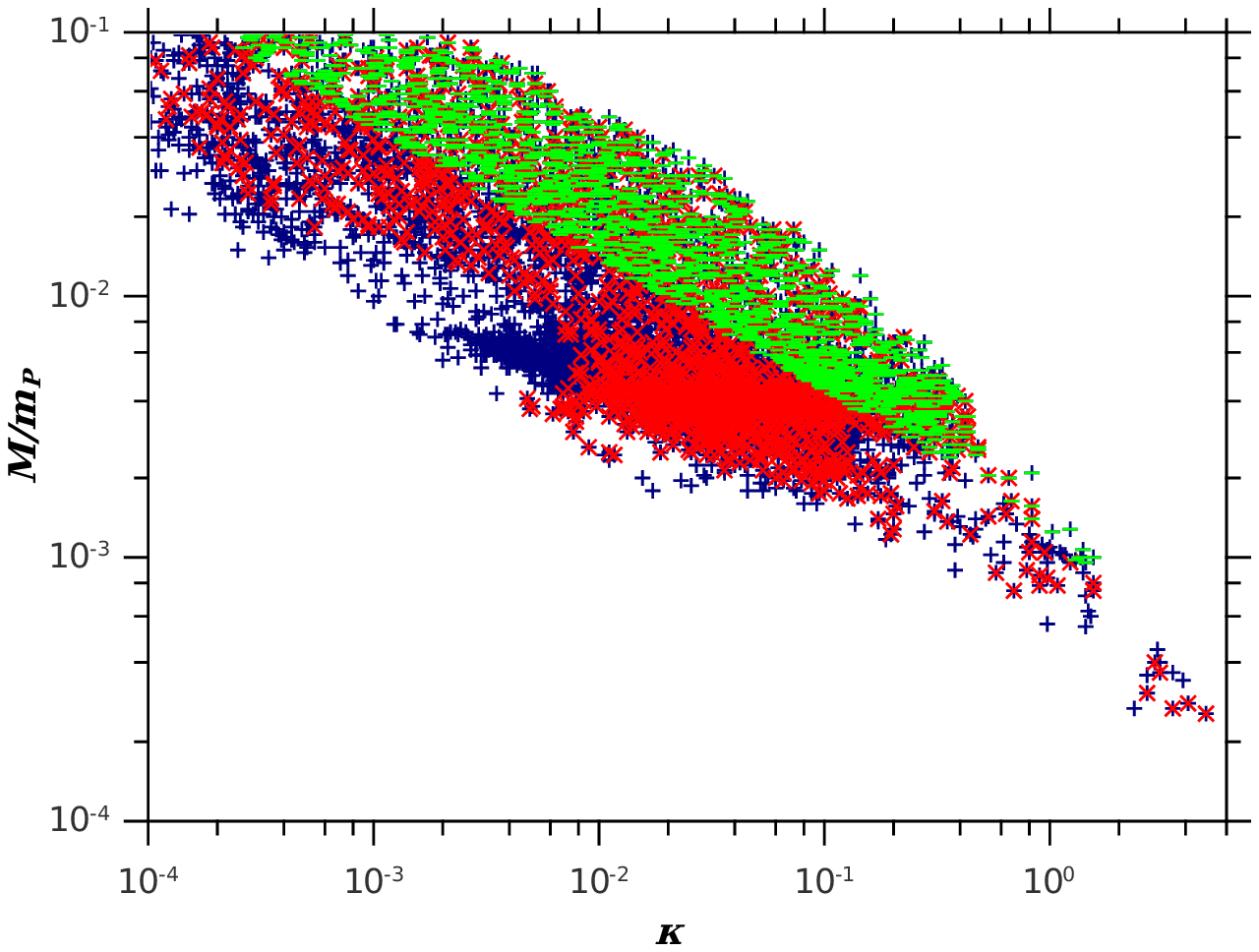} &
\includegraphics[width=6cm]{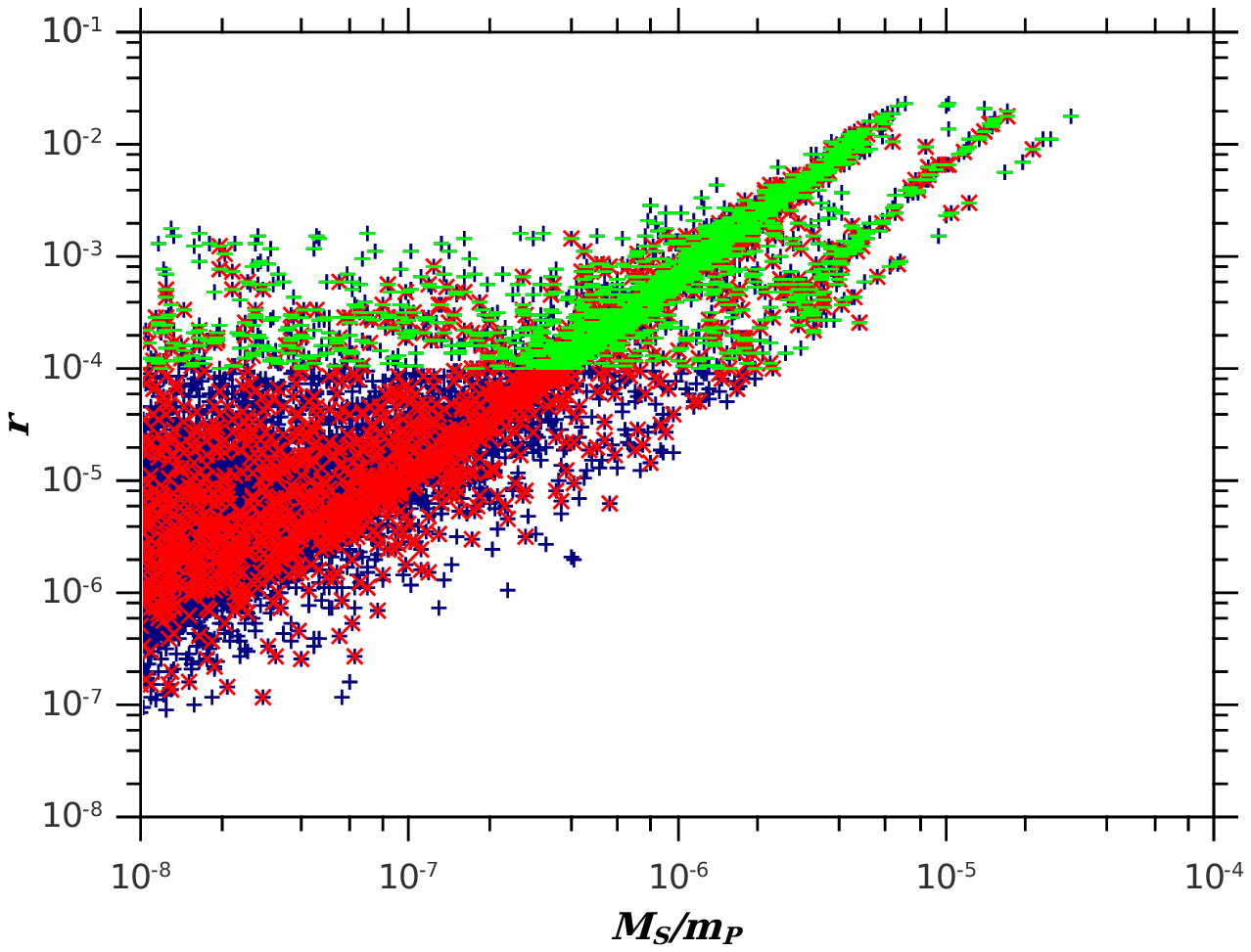} \\
(c) & (d) \\
\includegraphics[width=6cm]{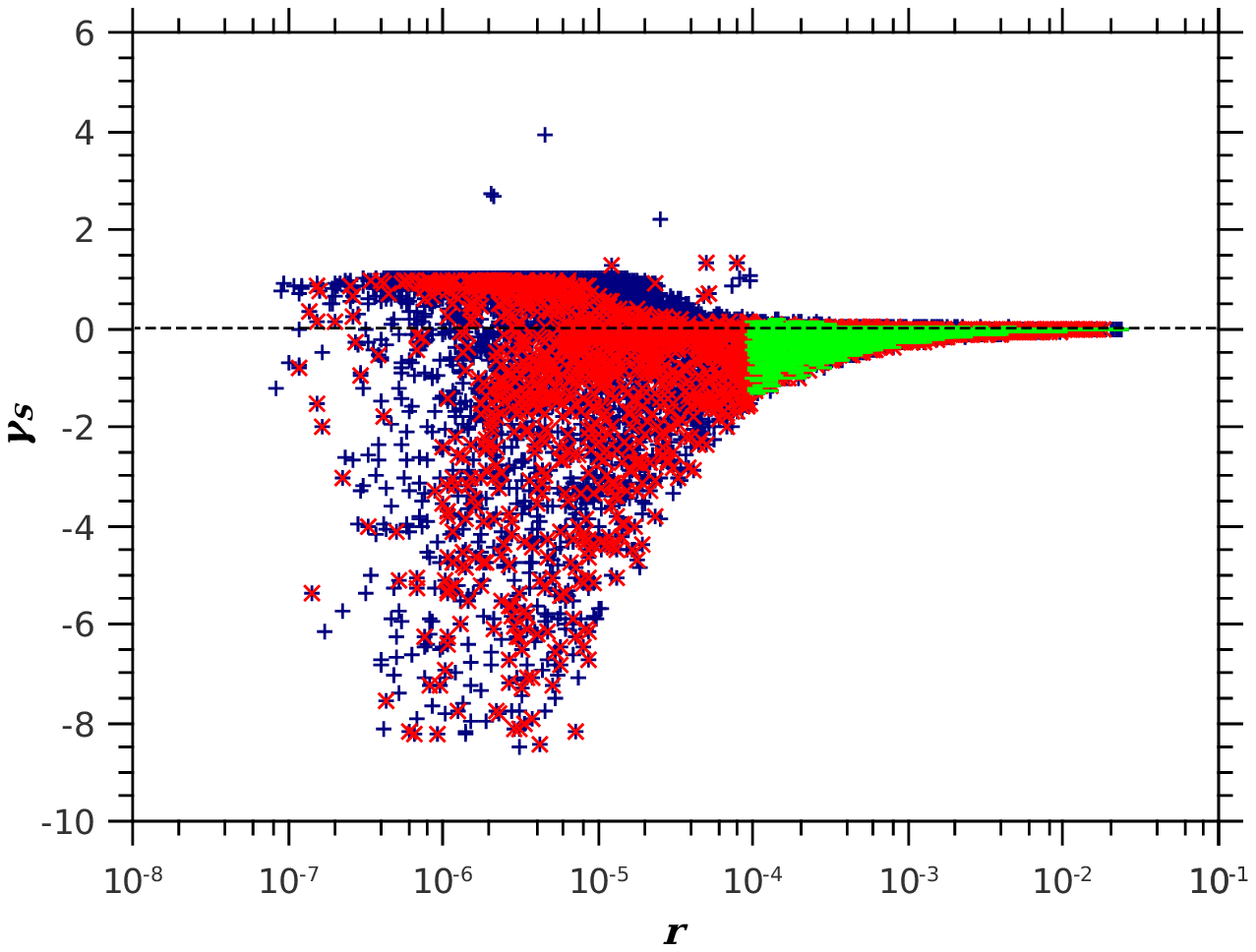} &
\includegraphics[width=6cm]{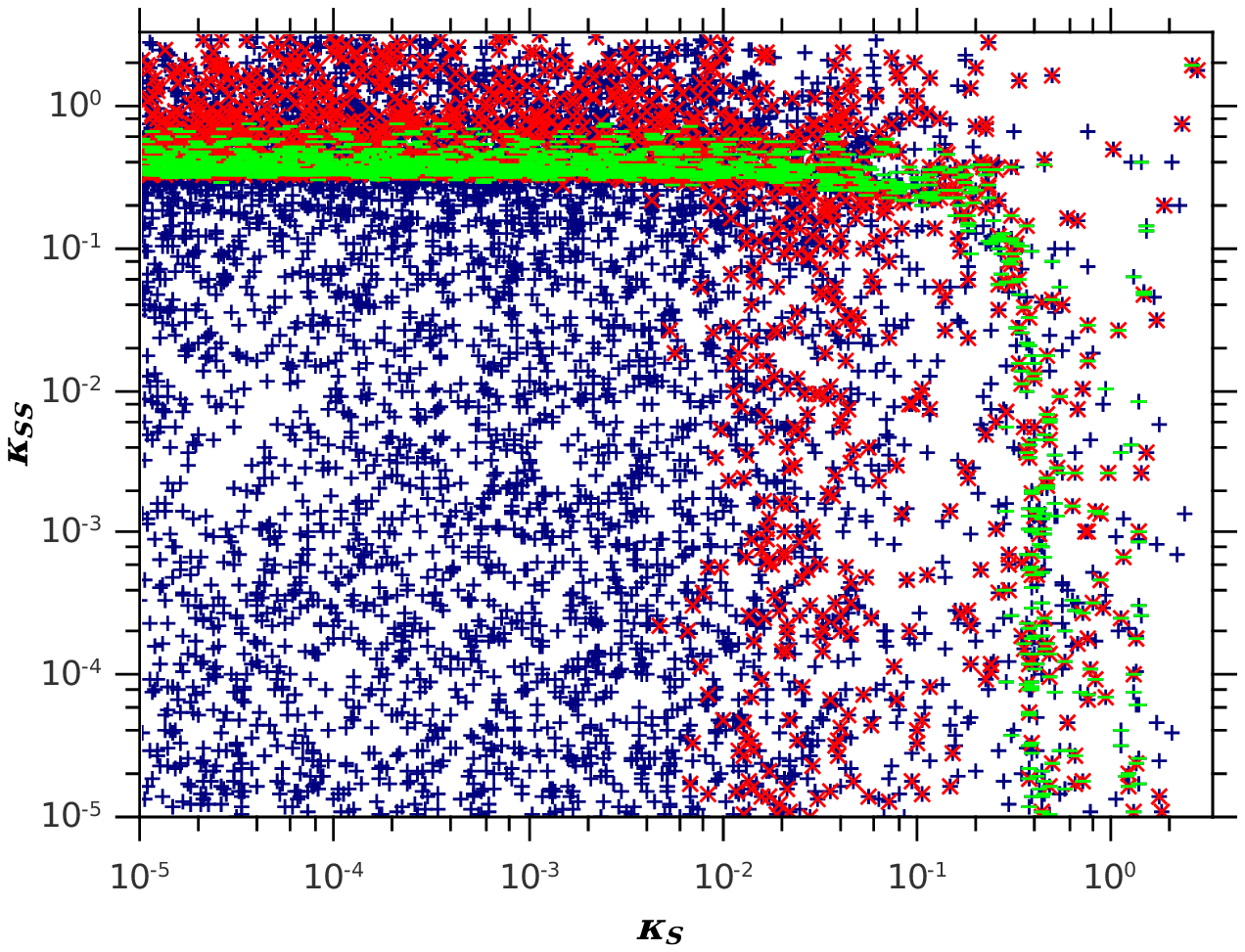} \\
(e) & (f) \\
\end{tabular}
\caption{Results of the numerical calculations for the flipped SU(5) case.  The points are color- and symbol-coded in the same fashion as Fig.~\ref{figpanel_NN1}.  While some parameters experience marginal shifts versus the U(1) case, the overall conclusions remain unchanged.}
\label{figpanel_NN10}
\end{figure}

%%%%%     flipped SU(5)

Finally, we turn our attention back to the issue of topological defect production.  Since the standard hybrid scenario offers no mechanism to suppress the density of defects produced (at the end of inflation) by the breaking of $G$, it may be advantageous to choose a gauge group whose breaking does not produce topological defects.  To this end, we also consider the so-called `flipped SU(5)' ($G\equiv \text{SU(5)} \times \text{U(1)}_X$) gauge group~\citep{Li:2010rz}, corresponding to $\mathcal{N} = 10$.  As an added bonus, it has recently been suggested that SUSY hybrid inflation models employing a flipped SU(5) GUT may lead to good predictions of the proton lifetime~\citep{Rehman:2009yj}.  As can be seen in Fig.~\ref{figpanel_NN10}, the change $\mathcal{N} = 1 \rightarrow 10$ produces small shifts in some parameters, but does not substantially affect our overall predictions.  Indeed, even the quantitative elements of the U(1) discussion above remain largely intact.  In particular, the flipped SU(5) case again leads to $r$-values up to $\approx 0.03$, within potential reach of the Planck satellite.

%%%%%%%%%%%%%%%%%%%%%%%%%%%%%%%%%%%%%%%%%%%%%%%%%

\section*{Summary}

Supersymmetric models of hybrid inflation, highly motivated by connections with mainstream particle physics, remain in good agreement with experiment.  We have shown that the typical predictions of the tensor-to-scalar ratio $r \ll 1$ can be assuaged via the use of a non-minimal K\"ahler potential in conjunction with a sizable soft inflaton mass.  We obtain values $r \lesssim 0.03$, which may be measurable by the current Planck satellite experiment.  This conclusion is made possible when the inflaton soft mass $M_S$ is quite large and the couplings associated with higher-order terms in the K\"ahler potential are allowed to have a significant magnitude.  (Note that there may exist solutions with $M_S \sim 1$~TeV if $\kappa_S < 0$, but this possibility remains to be explored.)  In addition, large tensor modes are obtained mainly for a concave-downward potential at the start of inflation reminiscent of hilltop inflation, which drives the spectral index $n_s$ toward red-tilted values.  The running of the spectral index, $|dn_s/d\ln k|$, can be $\lesssim 0.01$ in agreement with the latest experimental results.

%%%%%%%%%%%%%%%%%%%%%%%%%%%%%%%%%%%%%%%%%%%%%%%%%

\section*{Acknowledgments}

We thank Mansoor ur Rehman for many valuable discussions.  One of us (Q.S.) thanks Nefer {\c S}eno$\breve{\text{g}}$uz for helpful discussions.  This work is supported in part by the DOE under grant No.~DE-FG02-91ER40626, and by NASA and the Delaware Space Grant Consortium under grant No.\ NNG05GO92H (J.W.).

\bibliographystyle{h-elsevier}%{model1a-num-names}
\bibliography{hybridscanref}

\begin{thebibliography}{10}

\bibitem{Dvali:1994ms}
G.R. Dvali, Q. Shafi and R.K. Schaefer,
\newblock Phys. Rev. Lett. 73 (1994) 1886, hep-ph/9406319.

\bibitem{Copeland:1994vg}
E.J. Copeland et~al.,
\newblock Phys. Rev. D49 (1994) 6410, astro-ph/9401011.

\bibitem{Linde:1993cn}
A.D. Linde,
\newblock Phys. Rev. D49 (1994) 748, astro-ph/9307002.

\bibitem{Linde:1997sj}
A.D. Linde and A. Riotto,
\newblock Phys. Rev. D56 (1997) 1841, hep-ph/9703209.

\bibitem{Lazarides:2001zd}
G. Lazarides,
\newblock Lect. Notes Phys. 592 (2002) 351, hep-ph/0111328,
\newblock and references therein.

\bibitem{Lyth:1998xn}
D.H. Lyth and A. Riotto,
\newblock Phys. Rept. 314 (1999) 1, hep-ph/9807278.

\bibitem{Mazumdar:2010sa}
A. Mazumdar and J. Rocher,
\newblock (2010), 1001.0993,
\newblock for a recent review.

\bibitem{Jeannerot:2000sv}
R. Jeannerot et~al.,
\newblock JHEP 10 (2000) 012, hep-ph/0002151.

\bibitem{Senoguz:2003zw}
V.N. Senoguz and Q. Shafi,
\newblock Phys. Lett. B567 (2003) 79, hep-ph/0305089.

\bibitem{Kyae:2005nv}
B. Kyae and Q. Shafi,
\newblock Phys. Lett. B635 (2006) 247, hep-ph/0510105,
\newblock and references therein.

\bibitem{Lyth:1996im}
D.H. Lyth,
\newblock Phys. Rev. Lett. 78 (1997) 1861, hep-ph/9606387.

\bibitem{Lyth:2009zz}
D.H. Lyth and A.R. Liddle,
\newblock {The primordial density perturbation: Cosmology, inflation and the
  origin of structure} (Cambridge Univ. Pr., 2009).

\bibitem{Rehman:2009wv}
M.U. Rehman, Q. Shafi and J.R. Wickman,
\newblock Phys. Rev. D79 (2009) 103503, 0901.4345.

\bibitem{urRehman:2006hu}
M. ur~Rehman, V.N. Senoguz and Q. Shafi,
\newblock Phys. Rev. D75 (2007) 043522, hep-ph/0612023.

\bibitem{Rehman:2009nq}
M.U. Rehman, Q. Shafi and J.R. Wickman,
\newblock Phys. Lett. B683 (2010) 191, 0908.3896.

\bibitem{Rehman:2009yj}
M.U. Rehman, Q. Shafi and J.R. Wickman,
\newblock Phys. Lett. B688 (2010) 75, 0912.4737.

\bibitem{BasteroGil:2006cm}
M. Bastero-Gil, S.F. King and Q. Shafi,
\newblock Phys. Lett. B651 (2007) 345, hep-ph/0604198.

\bibitem{Clesse:2008pf}
S. Clesse and J. Rocher,
\newblock Phys. Rev. D79 (2009) 103507, 0809.4355.

\bibitem{Senoguz:2004vu}
V.N. Senoguz and Q. Shafi,
\newblock Phys. Rev. D71 (2005) 043514, hep-ph/0412102.

\bibitem{Jeannerot:2005mc}
R. Jeannerot and M. Postma,
\newblock JHEP 05 (2005) 071, hep-ph/0503146.

\bibitem{ArkaniHamed:2004fb}
N. Arkani-Hamed and S. Dimopoulos,
\newblock JHEP 06 (2005) 073, hep-th/0405159.

\bibitem{Giudice:2004tc}
G.F. Giudice and A. Romanino,
\newblock Nucl. Phys. B699 (2004) 65, hep-ph/0406088.

\bibitem{ArkaniHamed:2004yi}
N. Arkani-Hamed et~al.,
\newblock Nucl. Phys. B709 (2005) 3, hep-ph/0409232.

\bibitem{Komatsu:2010fb}
E. Komatsu et~al.,
\newblock (2010), 1001.4538,
\newblock It should be noted that, in the course of preparing this manuscript,
  a new version of the WMAP 7-year analysis paper was uploaded to the LAMBDA
  website, reflecting changes arising from the use of the latest versions of
  the analysis software. For our purposes, only the central value of $n_s$ has
  been modified; this new value has been used in the results that we present
  here.

\bibitem{Lin:2008ys}
C.M. Lin and K. Cheung,
\newblock JCAP 0903 (2009) 012, 0812.2731.

\bibitem{Vilenkin_book}
A. Vilenkin and E.P.S. Shellard,
\newblock Cosmic Strings and Other Topological Defects (Cambridge Univ. Pr.,
  1994),
\newblock and references therein.

\bibitem{Battye:2006pk}
R.A. Battye, B. Garbrecht and A. Moss,
\newblock JCAP 0609 (2006) 007, astro-ph/0607339.

\bibitem{Battye:2010xz}
R. Battye and A. Moss,
\newblock Phys. Rev. D82 (2010) 023521, 1005.0479.

\bibitem{Li:2010rz}
T. Li et~al.,
\newblock (2010), 1003.4186,
\newblock and references therein.

\bibitem{Boubekeur:2005zm}
L. Boubekeur and D.H. Lyth,
\newblock JCAP 0507 (2005) 010, hep-ph/0502047.

\bibitem{Kohri:2007gq}
K. Kohri, C.M. Lin and D.H. Lyth,
\newblock JCAP 0712 (2007) 004, 0707.3826.

\bibitem{Pallis:2009pq}
C. Pallis,
\newblock JCAP 0904 (2009) 024, 0902.0334.

\bibitem{Nakayama:2010xf}
K. Nakayama, F. Takahashi and T.T. Yanagida,
\newblock (2010), 1007.5152,
\newblock and references therein.

\bibitem{Lazarides:1985ja}
G. Lazarides, C. Panagiotakopoulos and Q. Shafi,
\newblock Phys. Rev. Lett. 56 (1986) 557.

\bibitem{Fukugita:1986hr}
M. Fukugita and T. Yanagida,
\newblock Phys. Lett. B174 (1986) 45.

\bibitem{Lazarides:1991wu}
G. Lazarides and Q. Shafi,
\newblock Phys. Lett. B258 (1991) 305,
\newblock for non-thermal leptogenesis.

\bibitem{Lazarides:1996dv}
G. Lazarides, R.K. Schaefer and Q. Shafi,
\newblock Phys. Rev. D56 (1997) 1324, hep-ph/9608256,
\newblock for leptogenesis in SUSY hybrid inflation.

\bibitem{Pilaftsis:2003gt}
A. Pilaftsis and T.E.J. Underwood,
\newblock Nucl. Phys. B692 (2004) 303, hep-ph/0309342.

\bibitem{DeSimone:2007pa}
A. De~Simone and A. Riotto,
\newblock JCAP 0708 (2007) 013, 0705.2183,
\newblock and references therein.

\bibitem{BenDayan:2009kv}
I. Ben-Dayan and R. Brustein,
\newblock JCAP 1009 (2010) 007, 0907.2384.

\end{thebibliography}

\end{document}